\documentclass[preprint]{aastex}

\usepackage{amsmath,amssymb,bm,mathrsfs}

\newcommand{\<}{{\kern-5pt}}

\newcommand{\thrj}[6]{\biggl(
	\arraycolsep .2em
	\begin{matrix}
	#1&#2&#3\\
	#4&#5&#6\\
	\end{matrix}\biggr)}
\newcommand{\sixj}[6]{\biggl\{
	\arraycolsep .2em
	\begin{matrix}
	#1&#2&#3\\
	#4&#5&#6\\
	\end{matrix}\biggr\}}

\def\mlangle{\kern.175em\langle}	
\def\mrangle{\rangle\kern.175em}	
\def\ket#1{|{#1}\mrangle}
\def\bra#1{\mlangle{#1}|}
\def\redmat#1#2#3{\bra{#1}\kern -1pt|#2|\kern -1pt\ket{#3}}

\newcommand{\apx}[1]{^{\mbox{\tiny{(#1)}}}}

\begin{document}

\title{Frequency Redistribution of Polarized Light\break 
in the $\Lambda$-Type Multi-Term Polarized Atom}

\author{R.\ Casini$^a$ and R.\ Manso Sainz$^b$}

\affil{$^a$High Altitude Observatory, National Center for Atmospheric
Research,\footnote{The National Center for Atmospheric Research is sponsored
by the National Science Foundation.}\break
P.O.~Box 3000, Boulder, CO 80307-3000, U.S.A.}
\affil{$^b$Max-Planck-Institut f\"ur Sonnensystemforschung,\break
Justus-von-Liebig-Weg 3, 37077 G\"ottingen, Germany}

\begin{abstract}
We study the effects of Rayleigh and Raman scattering on
the formation of polarized spectral lines in a $\Lambda$-type multi-term 
atom.
We fully take into account the partial redistribution of frequency 
and the presence of atomic
polarization in the lower states of the atomic model. Problems that can be modeled 
with this formalism include, for example, the formation of the \ion{Ca}{2} H-K and 
IR triplet, the analogous system of \ion{Ba}{2}, and the Ly$\beta$-H$\alpha$ system 
of hydrogenic ions.
\end{abstract}

\section{Introduction}

Many resonance lines of the solar spectrum show complex linear
polarization patterns, especially when observed near the solar limb
\citep{Wi75,SK96,SK97,St00,Ga02}.
These signals are generated by scattering in the upper photosphere and
chromosphere, where the plasma is very rarefied and nearly
collisionless, and long integration paths are involved in the production of
the observed polarized signals. Under these conditions, the ground state
of the observed transitions may harbour a significant amount of atomic
polarization \citep{TL97,TB02,MT03},
and coherence effects among the atomic levels become apparent 
in the emergent polarization profiles
\cite[e.g.,][]{St80,St97,La98,CM05,BT11,Sm12}

\begin{figure}[t!]
\centering
\includegraphics[width=.45\hsize]{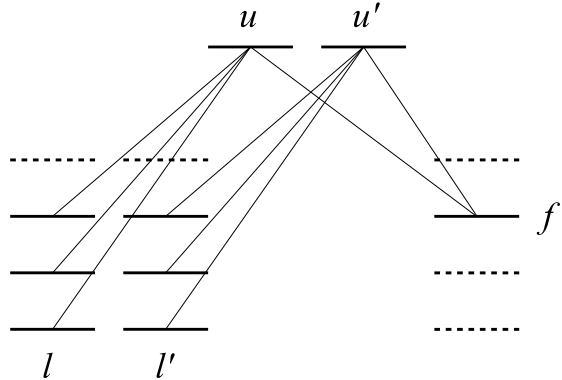}
\caption{Schematic diagram of the $\Lambda$-type, multi-term model atom 
considered in this 
paper, which is relevant for the modeling of the polarization of
the $u\to f$ transition. In order to correctly describe 
the polarization properties of the outgoing light, all lower terms $(l,l')$ 
that are radiatively connected to the (virtual) upper term $(u,u')$, also 
including the final term $f$, 
must be taken into account. Our model is restricted to the case where
all $(l,l')$ levels are sharp (i.e., with practically infinite lifetime).
\label{fig:model}}
\end{figure}

A general expression for the \emph{second-order emissivity} describing 
coherent\footnote{Following the terminology already adopted in 
Paper~I, here we use the term ``coherent'' in the broader sense of 
``memory preserving'', rather than in the traditional sense of 
``frequency preserving''. The latter meaning will instead be implied
when describing the results of Section~\ref{sec:application}.}
resonance scattering in a two-term polarized atom was recently 
presented by \citeauthor{Ca14} (\citeyear{Ca14}; hereafter, 
Paper~I).  On the other hand, very few lines in the solar 
spectrum can be considered strictly resonant. The ions in the solar 
atmosphere are illuminated by
a broadband radiation field that pumps the atomic levels through all 
possible transitions simultaneously, creating many different scattering 
channels. These conditions are responsible for the appearance of
complex coherence phenomena, where the scattered polarization 
signatures become coupled with the atomic polarization of 
the various energy levels involved in the atomic transitions.
The expression that was presented in Paper~I for the two-term atom
is general enough that it lends itself to a straightforward extension to 
the treatment of more complex atomic structures.
Here we generalize such expression to the description of polarized
radiation effects in a $\Lambda$-type multi-term atom, where 
the lower terms of the system are all radiatively coupled to a 
common excited state (see Figure~\ref{fig:model}).

An important example of such $\Lambda$-type system is contained in the 
\ion{Ca}{2} ion, and underlies the formation of some notable transitions 
observed in the solar spectrum: the H and K lines around 
395\,nm, which display a complex polarization pattern spanning more than
10\,nm \citep{St80,St80a,Ga02},
due to coherent scattering and quantum interference between the levels
of the upper ${}^2P$ term of the doublet \citep{St80};
and the infrared (IR) triplet around 858\,nm, in particular 
the 854.2\,nm and 866.2\,nm lines, which under solar conditions is 
dominated by the presence of atomic polarization in the metastable 
${}^2D$ term \citep{MT01,MT03}.
The UV doublet and the IR triplet are connected via the common 
${}^2P$ upper term, yet previous attemps at modeling the formation of
these sets of lines have not taken into account the full complexity of 
this system: either coherent scattering in a 
multi-level system was considered, but the essential polarization
of the metastable ${}^2D$ levels was neglected \citep{Sa13},
or the atomic polarization of all the levels was included in the
modeling of the IR triplet pattern, but coherence effects among the levels
of the upper ${}^2P$ term were neglected \citep{MT01,MT03}.


In the next Section, we provide explicit expressions for the 
$\Lambda$-type multi-term atom 
with and without hyperfine structure, as well as for the multi-level atom, 
when the fine-structure interaction is completely neglected. 
Finally, in 
Section~\ref{sec:application} we present the application of this formalism 
to the modeling of the polarized line profiles of some notable $\Lambda$-type
atomic systems.

\section{The $\Lambda$-type multi-term polarized atom with hyperfine 
structure}

We consider the general form of the radiative transfer equation for
polarized light in spectral lines, including the process of 
coherent scattering in a spectral line, which is responsible for the 
effects of partial redistribution of the radiation frequency when the 
illumination of the atom is not spectrally flat. Such transfer equation 
was presented in \cite{Ca14} (hereafter, Paper I), and when the polarized 
radiation is expressed in terms of the Stokes vector 
$\bm{S}\equiv(S_0,S_1,S_2,S_3)\equiv(I,Q,U,V)$, it takes the form
\begin{equation} \label{eq:RT.1}
\frac{1}{c}\,\frac{d}{dt}\,S_i(\omega_{k'},\bm{\hat{k}'}) =
	-\sum_j \kappa_{ij}(\omega_{k'},\bm{\hat{k}'})\,
		S_j(\omega_{k'},\bm{\hat{k}'})
	+\varepsilon\apx{1}_i(\omega_{k'},\bm{\hat{k}'})
	+\varepsilon\apx{2}_i(\omega_{k'},\bm{\hat{k}'})\;.\qquad (i=0,1,2,3)
\end{equation}
Here $\kappa_{ij}$ is the absorption matrix (corrected for 
radiation stimulated effects), and
$\varepsilon\apx{1}_i$ and $\varepsilon\apx{2}_i$ are the
source terms due to the spontaneous emission of radiation from
the excited levels, and to coherent scattering in the spectral 
line, respectively.

In the particular case of sharp lower levels, and in the absence of 
collisions, there is no true excitation of the upper levels, because of 
the infinite radiative lifetime of the lower levels. Then, the emissivity 
due to the spontaneous emission term $\varepsilon\apx{1}_i$ in the transfer 
equation can be neglected, as well as stimulation effects of the
radiation (see Paper I). 
In this case, the absorption term does not originate from true photon 
absorption, but corresponds instead to the atom's cross-section for 
the coherent scattering of radiation.

Accordingly, in this work we consider the scattering of radiation in 
resonance lines when both the initial and final states of the transition 
are infinitely sharp. Then the only \emph{positive} contribution to the line 
radiation comes from 
the coherent scattering term $\varepsilon\apx{2}_i$. This is given 
by Equation~(I.20), 
\begin{eqnarray} \label{eq:2emiss.1}
\varepsilon\apx{2}_i(\omega_{k'},\bm{\hat{k}'})
&\equiv&\frac{4}{3}\frac{e_0^4}{\hbar^2 c^4}\,{\cal N}\omega_{k'}^4
	\sum_{ll'}\rho_{ll'}\sum_{uu'f}
	\sum_{qq'}\sum_{pp'}(-1)^{q'+p'}\,
	(r_q)_{ul}(r_{q'})^\ast_{u'l'}
	(r_p)_{u'f}(r_{p'})^\ast_{uf} \\
&&\times
	\sum_{KQ}\sum_{K'Q'}\sqrt{(2K+1)(2K'+1)}\,
	\thrj{1}{1}{K}{-q}{q'}{-Q}
	\thrj{1}{1}{K'}{-p}{p'}{-Q'}\,
	T^{K'}_{Q'}(i,\bm{\hat{k}'}) \nonumber \\
&&\times \int_0^\infty d\omega_k
	\left(
	\Psi_{u'l',ful}^{-k,+k'-k} + \bar\Psi_{ul,fu'l'}^{-k,+k'-k}
	\right)
	J^K_Q(\omega_k)\;.\qquad (i=0,1,2,3) \nonumber
\end{eqnarray}
Here, $e_0(r_q)_{ab}$ represents the matrix element
between the atomic states $a$ and $b$ of the $q$ spherical component of
the electric dipole moment, $e_0\bm{r}$.
The geometric tensors $T^K_Q(i,\bm{\hat{k}})$ were introduced by
\cite{La84}, and their algebraic expressions have been tabulated
by several authors \citep[e.g.,][]{Bo97,LL04}. 
The radiation tensors $J^K_Q(\omega_k)$ are defined in terms of 
these geometric tensors and the incident Stokes vector as follows,
\begin{equation} \label{eq:Jtens}
J^K_Q(\omega_k)=\oint \frac{d\bm{\hat{k}}}{4\pi}\,
      \sum_{j=0}^3 T^K_Q(j,\bm{\hat{k}})\,S_j(\omega_k,\bm{\hat{k}})\;.
\end{equation}
Finally, the profiles $\Psi_{ab,cde}^{\pm h,\pm k\pm l}$, which describe
the effects of frequency redistribution, are given by Equation~(I.5). 
We indicate the complex conjugate of these profiles with
$\bar\Psi_{ab,cde}^{\pm h,\pm k\pm l}$.

Equation~(\ref{eq:2emiss.1}) describes the scattering of polarized 
light that occurs in the radiative transition from a set of lower 
levels $(l,l')$ -- weighted by the density matrix $\rho_{ll'}$ and the
gas density $\cal N$ for the ionic species considered -- to a 
final set of levels $f$, via the virtual excitation of a set of 
intermediate upper levels $(u,u')$. It is important to observe
that this expression generally describes the scattering polarization of
a radiative transition $u\to f$ that results from the combination of 
distinct $\Lambda$-type atomic systems sharing the same final branch 
$(u,u')\to f$, but with different initial branches $(l,l')\to (u,u')$ 
(see Figure~\ref{fig:model}).
Therefore, Equation~(\ref{eq:2emiss.1}) can be used to model
the scattering of polarized radiation in multi-term atoms of the 
$\Lambda$-type, including the effects of a magnetic field and of
lower-level polarization. We consider the general case where the hyperfine 
structure may be present, since such a model is necessary to 
describe many interesting spectral lines of the solar chromosphere
(see Table~\ref{tab:lines}).

We must observe that the index substitution $l''\to f$ in the original 
expression (I.20) for the second-order emissivity of a two-term atom, 
which allowed us to write equation (2), is not just a formal exercise. 
There are supporting physical arguments showing that this new expression 
actually applies to the case of a $\Lambda$-type multi-term atom. For 
example, it reproduces the spectral emission of such a system when 
the incident radiation field is spectrally flat, i.e., in the limit of 
complete redistribution of the incident energy (see Paper \ I, Sect. 6). 
This correct behavior was numerically verified in all atomic models that 
we tested.

\begin{table}[!t]
\centering
\caption{\label{tab:lines}
Notable $\Lambda$-type multiplets of the polarized solar spectrum 
observed near the limb \citep[see, e.g.][]{St83a,St83b,Ga00,Ga02,Ga05}}
\vskip 6pt 
\begin{tabular}{cccc}
\hline\hline 
Ion & $\lambda$ (nm) & lower terms & upper term \\
\noalign{\vskip 2pt}
\hline \noalign{\vskip 2pt}  
\ion{H}{1}    & $\rm 102.5\;(Ly\beta)$      &  $1s\; {}^2\!S$ &  $3p\;  {}^2\!P^\circ$ \\
             &  $\rm 656.3\;(H\alpha)$      &  $2s\; {}^2\!S$ &  \\
\hline \noalign{\vskip 2pt}  
\ion{Ca}{2}  &  $\rm 393.4\;(K),396.8\;(H)$    &  $4s\; {}^2\!S$ &  $4p\; {}^2\!P^\circ$ \\
             &  $849.8,854.2,866.2$   &  $3d\; {}^2\!D$ &   \\
\noalign{\vskip 2pt}
\hline \noalign{\vskip 2pt}  
\ion{Ba}{2}  &  $\rm 455.4\;(D_2),493.4\;(D_1)$ & $6s\; {}^2\!S$ &  $6p\; {}^2\!P^\circ$ \\
             &  $585.4,614.2,649.7$      &  $5d\; {}^2\!D$ &   \\
\noalign{\vskip 2pt} 
\hline \noalign{\vskip 2pt}  
\ion{Sc}{2}  & $424.7$    &   $3p^63d4s\; {}^1\!D$  &    $3p^63d4p \;{}^1\!D^\circ$ \\
             & $660.5$    &   $3p^63d^2\; {}^1\!D$  &  \\
\noalign{\vskip 2pt} 
\hline \noalign{\vskip 2pt}  
\ion{Cr}{1}   & $520.4,520.6,520.8$ &   $a\; {}^5\!S$  &$z \;{}^5\!P^\circ$  \\
             & $524.7,\ldots,540.9$     &   $a\; {}^5\!D$  &     \\
             & 2017.9\dots2024.6   &   $a\; {}^5\!P$  &   \\
\noalign{\vskip 2pt} 
\hline\hline  
\end{tabular}
\end{table}

We indicate with $\alpha_l$ and $\alpha_{f}$ the electronic
configuration of the two lower terms representing respectively the initial 
and final states of the transition, and correspondingly with $\alpha_u$ 
the electronic configuration of the intermediate upper term (see 
Figure~\ref{fig:model}). 
We assume the direction of the magnetic field as the quantization axis 
($z$-axis). Then the atomic states involved in Equation~(\ref{eq:RT.1}) 
are of the form
\begin{eqnarray*}
l&\equiv&\alpha_l I\mu_l M_l\;,\quad
l'\equiv\alpha_l I\mu_l' M_l'\;,\\
u&\equiv&\alpha_u I\mu_u M_u\;,\quad
u'\equiv\alpha_u I\mu_u' M_u'\;,\\
f&\equiv&\alpha_{f} I\mu_f M_f\;,
\end{eqnarray*}
where $M$ is the projection of the total angular momentum $\bm{F}$ on
the $z$-axis, $I$ is the quantum number of the nuclear spin, while 
$\mu$ is the index of the atomic Hamiltonian eigenbasis spanning the 
subspace of all the quantum numbers $J$ and $F$ that are associated with 
a given value of $M$.

Following the formalism of Paper~I, Equation~(\ref{eq:2emiss.1}) becomes
\begin{eqnarray} \label{eq:RT.JF}
\varepsilon\apx{2}_i(\omega_{k'},\bm{\hat{k}'})
&=& \frac{3}{16\pi^3}\,
	{\cal N}\hbar\,\omega_{k'}^4\,
	\Pi_{L_u}^2
	\frac{A_{uf}}{\omega_{uf}^3}
	\sum_{L_l} \Pi_{L_l}^2 B_{lu}\, \\
&&\kern -1.5cm\times
	\sum_{J_u J_u' J_u'' J_u'''}
	\sum_{J_l J_l' J_f J_f'}
	\sum_{F_u F_u' F_u'' F_u'''}
	\sum_{F_l F_l' F_f F_f'}
	(-1)^{J_u+J_u'+J_u''+J_u'''}
	(-1)^{J_l+J_l'+J_f+J_f'}
	\nonumber \\
&&\mathop{\times}
	\Pi_{J_u J_u' J_u'' J_u'''}\,
	\Pi_{J_l J_l' J_f J_f'}\,
	\sixj{J_u}{J_l}{1}{L_l}{L_u}{S}
	\sixj{J_u'}{J_l'}{1}{L_l}{L_u}{S}
	\sixj{J_u''}{J_f}{1}{L_{f}}{L_u}{S}
	\sixj{J_u'''}{J_f'}{1}{L_{f}}{L_u}{S} \nonumber \\
&&\mathop{\times}
	\Pi_{F_u F_u' F_u'' F_u'''}\,
	\Pi_{F_l F_l' F_f F_f'}\,
	\sixj{F_u}{F_l}{1}{J_l}{J_u}{I}
	\sixj{F_u'}{F_l'}{1}{J_l'}{J_u'}{I}
	\sixj{F_u''}{F_f}{1}{J_f}{J_u''}{I}
	\sixj{F_u'''}{F_f'}{1}{J_f'}{J_u'''}{I} \nonumber \\
&&\kern -1.5cm \times
	\sum_{\mu_u M_u}\sum_{\mu_u' M_u'}\sum_{\mu_f M_f}
	C^{J_u F_u}_{\mu_u}(M_u)\,
	C^{J_u'' F_u''}_{\mu_u}(M_u)\,
	C^{J_u' F_u'}_{\mu_u'}(M_u')\,
	C^{J_u''' F_u'''}_{\mu_u'}(M_u')\,
	C^{J_f F_f}_{\mu_f}(M_f)\,
	C^{J_f' F_f'}_{\mu_f}(M_f) \nonumber \\
&&\kern -1.5cm \times 
	(\epsilon_{uu'}+{\rm i} \omega_{uu'})^{-1}
	\sum_{\bar J_l \bar J_l'}
	\sum_{\bar F_l \bar F_l'}
	\sum_{\mu_l M_l}\sum_{\mu_l' M_l'}
	C^{J_l F_l}_{\mu_l}(M_l)\,
	C^{\bar J_l \bar F_l}_{\mu_l}(M_l)\,
	C^{J_l' F_l'}_{\mu_l'}(M_l')\,
	C^{\bar J_l' \bar F_l'}_{\mu_l'}(M_l') \nonumber \\
&&\kern -1.5cm \times
	\sum_{KQ}\sum_{K'Q'}\sum_{K_l Q_l}
	\sum_{qq'} \sum_{pp'}
	(-1)^{\bar F_l-M_l+q'+p'}
	\thrj{1}{1}{K}{-q}{q'}{-Q}
	\thrj{1}{1}{K'}{-p}{p'}{-Q'}
	\thrj{\bar F_l}{\bar F_l'}{K_l}{M_l}{-M_l'}{-Q_l}
	\nonumber \\
&&\mathop{\times}
	\thrj{F_u}{F_l}{1}{-M_u}{M_l}{q}
	\thrj{F_u'}{F_l'}{1}{-M_u'}{M_l'}{q'}
	\thrj{F_u'''}{F_f'}{1}{-M_u'}{M_f}{p}
	\thrj{F_u''}{F_f}{1}{-M_u}{M_f}{p'}
	\nonumber \\
&&\mathop{\times}
	\Pi_{KK'K_l}\,
	T^{K'}_{Q'}(i,\bm{\hat{k}'})\,
	\rho^{K_l}_{Q_l}(\bar J_l \bar F_l,\bar J_l' \bar F_l') 
	\nonumber \\
&&\kern -1.5cm\times 
      \sum_{j=0}^3 \oint \frac{d\bm{\hat{k}}}{4\pi}\,
	T^K_Q(j,\bm{\hat{k}})
	\int_0^\infty d\omega_k\;
	{\cal R}(\Omega_u,\Omega_{u'};
	\Omega_l,\Omega_{l'},\Omega_f;
	\omega_k,\omega_{k'})\,
	S_j(\omega_k,\bm{\hat{k}})\;,\qquad (i=0,1,2,3)
	\nonumber
\end{eqnarray}
where we adopted the shorthand notation
%
$\Pi_{ab\ldots}\equiv\sqrt{(2a+1)(2b+1)\cdots}$.
%
The various coefficients $C_\mu^{JF}(M)$ represent the projection 
components of the eigenstates $\ket{\mu M}$ of the magnetic Hamiltonian
on the basis of atomic states of the form $\ket{(JI)FM}$. In writing 
equation~(\ref{eq:RT.JF}) we also introduced the redistribution function
in the atomic frame of reference,
\begin{equation} \label{eq:redist}
{\cal R}(\Omega_u,\Omega_{u'};
	\Omega_l,\Omega_{l'},\Omega_{l''};
	\omega_k,\omega_{k'})
	\equiv  
	(\epsilon_{uu'}+{\rm i}\omega_{uu'})
	\left(
	\Psi_{u'l',l''ul}^{-k,+k'-k} + \bar\Psi_{ul,l''u'l'}^{-k,+k'-k}
	\right)\;.
\end{equation}

The transformation of equation~(\ref{eq:RT.JF}) to the laboratory frame
of reference is formally attained by replacing the redistribution
function of equation~(\ref{eq:redist}) with the appropriate
velocity-dependent function 
$R(\Omega_u,\Omega_{u'};\Omega_l,\Omega_{l'},\Omega_{l''};
	\hat{\omega}_{k},\hat{\omega}_{k'};\Theta)$,
where $\hat{\omega}_{k}$ and $\hat{\omega}_{k'}$ are the frequencies 
of the incoming and outgoing radiation in the laboratory
frame, respectively, and $\Theta$ is the scattering angle.
The task of extending the redistribution function in the laboratory
frame to the case of a $\Lambda$-type three-term polarized atom has 
been undertaken in a separate work (R.\ Casini \& R.\ Manso 
Sainz 2016; in preparation). 
In the Appendix, we give the form of such redistribution function 
in the limit case of \emph{non-coherent lower term}, which applies 
practically to all the examples presented in 
Section~\ref{sec:application}.

It is important to note that the widths of the level $u$ and $u'$ 
appearing in the redistribution profiles of the form
$\Psi_{u'l',ful}^{-k,+k'-k}$ must take into account all possible 
spontaneous de-excitation processes towards lower terms $l$, i.e.,
\begin{equation}  \label{eq:widths}
\epsilon_u=\frac{1}{2}\sum_l A_{ul}\approx\epsilon_{u'}\;,
\end{equation}
where, according to our model, the set of lower terms $l$ also 
includes the final term $f$.

For some applications, including those presented in the next section,
it is necessary to consider atomic models without hyperfine 
structure, i.e., $I=0$. In that case, 
after some straightforward Racah algebra manipulations of
equation~(\ref{eq:RT.JF}), we find
\begin{eqnarray} \label{eq:RT.J}
\varepsilon\apx{2}_i(\omega_{k'},\bm{\hat{k}'})
&=& \frac{3}{16\pi^3}\,
	{\cal N}\hbar\,\omega_{k'}^4\,
	\Pi_{L_u}^2 
	\frac{A_{uf}}{\omega_{uf}^3}
	\sum_{L_l} \Pi_{L_l}^2 B_{lu} \\ 
&&\kern -1.5cm\times
	\sum_{J_u J_u' J_u'' J_u'''}
	\sum_{J_l J_l' J_f J_f'}
	\Pi_{J_u J_u' J_u'' J_u'''}\,
	\Pi_{J_l J_l' J_f J_f'}\,
	\sixj{J_u}{J_l}{1}{L_l}{L_u}{S}
	\sixj{J_u'}{J_l'}{1}{L_l}{L_u}{S}
	\sixj{J_u''}{J_f}{1}{L_{f}}{L_u}{S}
	\sixj{J_u'''}{J_f'}{1}{L_{f}}{L_u}{S} \nonumber \\
&&\kern -1.5cm \times
	\sum_{\mu_u M_u}\sum_{\mu_u' M_u'}\sum_{\mu_f M_f}
	C^{J_u}_{\mu_u}(M_u)\,
	C^{J_u''}_{\mu_u}(M_u)\,
	C^{J_u'}_{\mu_u'}(M_u')\,
	C^{J_u'''}_{\mu_u'}(M_u')\,
	C^{J_f}_{\mu_f}(M_f)\,
	C^{J_f'}_{\mu_f}(M_f) \nonumber \\
&&\kern -1.5cm \times 
	(\epsilon_{uu'}+{\rm i} \omega_{uu'})^{-1}
	\sum_{\bar J_l \bar J_l'}
	\sum_{\mu_l M_l}\sum_{\mu_l' M_l'}
	C^{J_l}_{\mu_l}(M_l)\,
	C^{\bar J_l}_{\mu_l}(M_l)\,
	C^{J_l'}_{\mu_l'}(M_l')\,
	C^{\bar J_l'}_{\mu_l'}(M_l') \nonumber \\
&&\kern -1.5cm \times
	\sum_{KQ}\sum_{K'Q'}\sum_{K_l Q_l}
	\sum_{qq'} \sum_{pp'}
	(-1)^{\bar J_l-M_l+q'+p'}
	\thrj{1}{1}{K}{-q}{q'}{-Q}
	\thrj{1}{1}{K'}{-p}{p'}{-Q'}
	\thrj{\bar J_l}{\bar J_l'}{K_l}{M_l}{-M_l'}{-Q_l}
	\nonumber \\
&&\mathop{\times}
	\thrj{J_u}{J_l}{1}{-M_u}{M_l}{q}
	\thrj{J_u'}{J_l'}{1}{-M_u'}{M_l'}{q'}
	\thrj{J_u'''}{J_f'}{1}{-M_u'}{M_f}{p}
	\thrj{J_u''}{J_f}{1}{-M_u}{M_f}{p'}
	\nonumber \\
&&\mathop{\times}
	\Pi_{KK'K_l}\,
	T^{K'}_{Q'}(i,\bm{\hat{k}'})\,
	\rho^{K_l}_{Q_l}(\bar J_l,\bar J_l') 
	\nonumber \\
&&\kern -1.5cm\times 
      \sum_{j=0}^3 \oint \frac{d\bm{\hat{k}}}{4\pi}\,
	T^K_Q(j,\bm{\hat{k}})
	\int_0^\infty d\omega_k\;
	{\cal R}(\Omega_u,\Omega_{u'};
	\Omega_l,\Omega_{l'},\Omega_f;
	\omega_k,\omega_{k'})\,
	S_j(\omega_k,\bm{\hat{k}})\;,\qquad (i=0,1,2,3)
	\nonumber
\end{eqnarray}
%


%

Finally, in the case of the multi-level atom, there is no dependence 
of the line profiles on the $\mu$-indices, and so we can use the 
orthogonality properties of the Hamiltonian eigenvectors (see 
equations~(I.23)) in order to perform the trivial summations over 
those indices. The expression of the second-order emissivity for this
model atom can be derived directly from equation~(\ref{eq:RT.J})
by imposing the additional conditions $S=0$ and $L=J$: 
\begin{eqnarray} \label{eq:RT.level}
\varepsilon\apx{2}_i(\omega_{k'},\bm{\hat{k}'})
&=& \frac{3}{16\pi^3}\,
	{\cal N}\hbar\,\omega_{k'}^4\,
	\Pi_{J_u}^2
	\frac{A_{uf}}{\omega_{uf}^3}
	\sum_{J_l} \Pi_{J_l}^2 B_{lu} \\
&&\kern -1.5cm \times
	\sum_{M_u M_u'}
	\sum_{M_l M_l'}\sum_{M_f}
	\sum_{KQ}\sum_{K'Q'}\sum_{K_l Q_l}
	\sum_{qq'} \sum_{pp'}
	(-1)^{J_l-M_l+q'+p'}
	\thrj{1}{1}{K}{-q}{q'}{-Q}
	\thrj{1}{1}{K'}{-p}{p'}{-Q'}
	\nonumber \\
&&\mathop{\times}
	\thrj{J_u}{J_l}{1}{-M_u}{M_l}{q}
	\thrj{J_u}{J_l}{1}{-M_u'}{M_l'}{q'}
	\thrj{J_u}{J_f}{1}{-M_u'}{M_f}{p}
	\thrj{J_u}{J_f}{1}{-M_u}{M_f}{p'}
	\thrj{J_l}{J_l}{K_l}{M_l}{-M_l'}{-Q_l}
	\nonumber \\
&&\mathop{\times}
	(\epsilon_{uu'}+{\rm i} \omega_{uu'})^{-1}\,
	\Pi_{KK'K_l}\,
	T^{K'}_{Q'}(i,\bm{\hat{k}'})\,
	\rho^{K_l}_{Q_l}(J_l) 
	\nonumber \\
\noalign{\allowbreak}
&&\kern -1.5cm\times 
      \sum_{j=0}^3 \oint \frac{d\bm{\hat{k}}}{4\pi}\,
	T^K_Q(j,\bm{\hat{k}})
	\int_0^\infty d\omega_k\;
	{\cal R}(\Omega_u,\Omega_{u'};
	\Omega_l,\Omega_{l'},\Omega_f;
	\omega_k,\omega_{k'})\,
	S_j(\omega_k,\bm{\hat{k}})\;.\qquad (i=0,1,2,3)
	\nonumber
\end{eqnarray}
%

In the next section, we provide examples of the application of this
formalism (specifically, of equation~(\ref{eq:RT.J})) to a few notable
$\Lambda$-type three-term atoms, namely
the Ly$\beta$-H$\alpha$ system of hydrogen, and the \ion{Ca}{2} 
H-K doublet with the IR triplet (see Table~\ref{tab:lines}).

\begin{figure}[t!]
\centering
\includegraphics[width=.85\hsize]{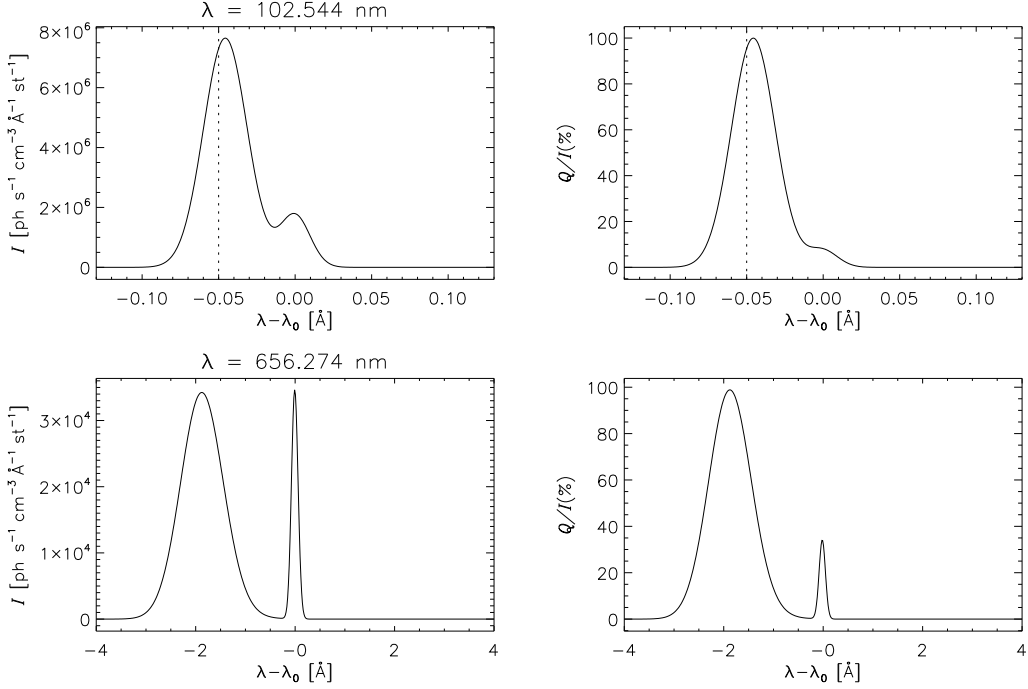}
\caption{\label{fig:Lyb-Ha}
Stokes $I$ and $Q$ emission profiles of the \ion{H}{1} Ly$\beta$ (top panels) 
and H$\alpha$ (bottom panels) lines, respectively at 102.5\,nm and 
656.3\,nm. The model is for the 90-degree scattering of a collimated 
beam of radiation by a plasma with temperature $T=1000$\,K, and 
no magnetic field. The intensity of the incident radiation field 
corresponds to a Planckian spectrum at $T_{\rm rad}=20\,000$\,K.
The incident UV radiation (monochromatic and unpolarized) 
is nearly resonant with the Ly$\beta$ transition, with a detuning 
of $-0.05$\,\AA\ (vertical dashed lines in the top panels). We note 
the presence of a secondary contribution in the scattered radiation,
which is perfectly resonant with the natural wavelength of the line.
The intensity profile amplitude is expressed as number of scattered
photons per unit intervals of time, emitting volume, wavelength, and 
solid angle, assuming a reference gas density 
${\cal N}=10^{12}\,\rm cm^{-3}$ for the emitting volume.} 
\end{figure}

\section{Examples of partial redistribution in $\Lambda$-type 
three-term polarized atoms}
\label{sec:application}

As an application of equation~(\ref{eq:RT.J}), we first consider the 
simplest case of the $\Lambda$-type system $1s$-$3p$-$2s$ of \ion{H}{1}, 
which pertains to the formation of the Ly$\beta$
and H$\alpha$ lines, respectively at 102.5\,nm and 656.3\,nm.
Since both lines formed in this restricted $\Lambda$-type system
have sharp lower levels, the corresponding three-term model atom can 
indeed be described through the formalism presented above.
More specifically, we model the effect of detuning of the
UV radiation around the wavelength of the Ly$\beta$ line on the intensity 
and polarization of the H$\alpha$ line. 
For this purpose, we assume an ensemble of \ion{H}{1} atoms
with a pre-assigned distribution of population and atomic polarization
in the ground and metastable states. This distribution is derived by 
solving the statistical equilibrium for the atomic 
system under prescribed illumination conditions, corresponding to a 
collimated beam of highly diluted radiation with a Planckian spectrum
at $T_\textrm{rad}=20\,000\,\rm K$. Under such conditions, the 
population of the excited $3p\,{}^2P^\circ$ term is negligible with respect 
to the populations of the ground and metastable states (by about 9 and 8 
orders of magnitude, respectively), and the PRD formalism 
presented in this work, where the incoherent emission of photons from
spontaneous de-excitation of the upper term can be neglected, thus 
becomes applicable. We also remark that the atomic system so
prepared will in general harbor atomic polarization, because of the
condition of anisotropic illumination associated with the collimated beam
of incident radiation.

\begin{figure}[t!]
\centering
\includegraphics[width=\hsize]{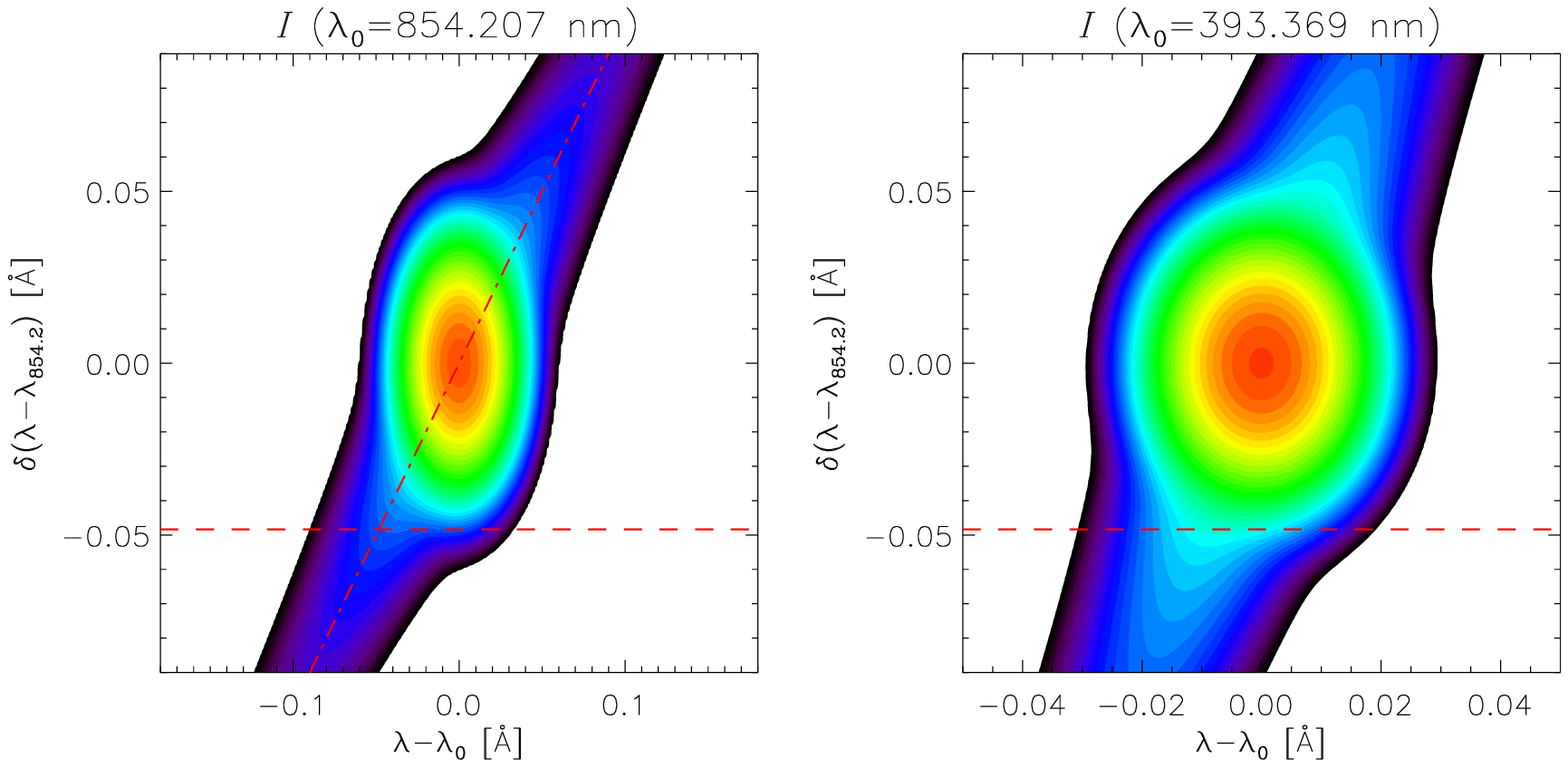}
\caption{\label{fig:plots.res.8542}
Contour plots of the scattered intensity of the IR line at 854.2\,nm (left) 
and the K line at 393.4\,nm (right) of \ion{Ca}{2}, as a function of 
wavelength across the line spectral ranges ($x$ axes), and of the 
detuning of the incident monochromatic IR radiation around the 
854.2\,nm resonance
wavelength ($y$ axis). The scattering configuration is the same
as for Figure~\ref{fig:Lyb-Ha}, but for an incident radiation with
a Planckian spectrum
at $T_{\rm rad}=5000$\,K. The diagonal 
dash-dotted line in the left panel tracks the location of the
monochromatic illumination with respect to the intensity profile of the
scattered 854.2\,nm line, for the various values of detuning.
The horizontal dashed lines locate the value of the detuning of the 
monochromatic IR incident radiation adopted for the calculation of the 
emission profiles shown in Figure~\ref{fig:plots.offres.8542}.}
\end{figure}

Figure~\ref{fig:Lyb-Ha} shows the scattered radiation produced in this 
system. The wavelength of the UV incident radiation, which again
is monochromatic and unpolarized, is marked by the vertical dashed lines 
in the Stokes $I$ and $Q$ plots of the Ly$\beta$ line.
In order to clearly identify the various contributions to the
scattered radiation in the system, we assume no incident radiation at 
the H$\alpha$ wavelength. Hence, the
radiation scattered in the spectral range of H$\alpha$ is completely due
to Raman scattering through virtual excitation of the $3p\,{}^2P^\circ$ 
term by the UV radiation.
%

We note that the scattered radiation in the Ly$\beta$ line is dominated by 
the coherent component centered around the wavelength of the
monochromatic UV incident radiation. However, because of the relatively 
large Doppler width compared to the size of the detuning, a small 
contribution, resonant with the natural transition of the line 
and blended with the coherent component, is also present in 
this case. In contrast, the emission in H$\alpha$ shows these same 
two contributions well separated. In fact, if we indicate with 
$\delta\lambda_{ul}$ the detuning of the incident radiation in the 
$\Lambda$-type system, for the final branch of the scattering process
\begin{equation} \label{eq:important}
\delta\lambda_{uf}=\left(\frac{\lambda_{uf}}{\lambda_{ul}}\right)^2
	\delta\lambda_{ul}\;.
\end{equation}
Because in Figure~\ref{fig:Lyb-Ha} the detuning from the Ly$\beta$ 
resonance is $\delta\lambda_{ul}=-0.05$\,\AA, the coherent component 
in H$\alpha$ occurs with a wavelength shift from resonance given by 
$\delta\lambda_{uf}\approx -2.05$\,\AA. As this is larger than the 
Doppler width corresponding to the assumed plasma temperature (which 
instead scales linearly with the wavelength of the transition), the 
coherent and resonant contributions in the H$\alpha$ line appear 
completely separated.
We also note that the line profiles of both transitions are 100\% linearly 
polarized at the frequency of coherent re-emission, as it is to be expected 
in the wings of $S$-$P$ transitions.
Finally, we point out that the wavelength integrated intensity 
profiles of Figure~\ref{fig:Lyb-Ha}, which give the total numbers of 
photons emitted in the Ly$\beta$ and H$\alpha$ lines, are in exactly the 
same ratio as the Einstein A-coefficients of the two transitions, which
is also to be expected.

Next, we consider the more complex atomic system of \ion{Ca}{2}, leading to 
the formation of the H and K lines around 395\,nm and the IR triplet around 
858\,nm. 
We model the effect of detuning in the two cases where the
monochromatic incident radiation 
is located in the proximity of either one or the other of the two 
transitions.
We consider the same scattering configuration as in the previous example,
with the exception of the temperature of the incident radiation, 
for which we assume instead a Planckian spectrum at
$T_\textrm{rad}=5000\,\rm K$. 

\begin{figure}[t!]
\centering
\includegraphics[width=.85\hsize]{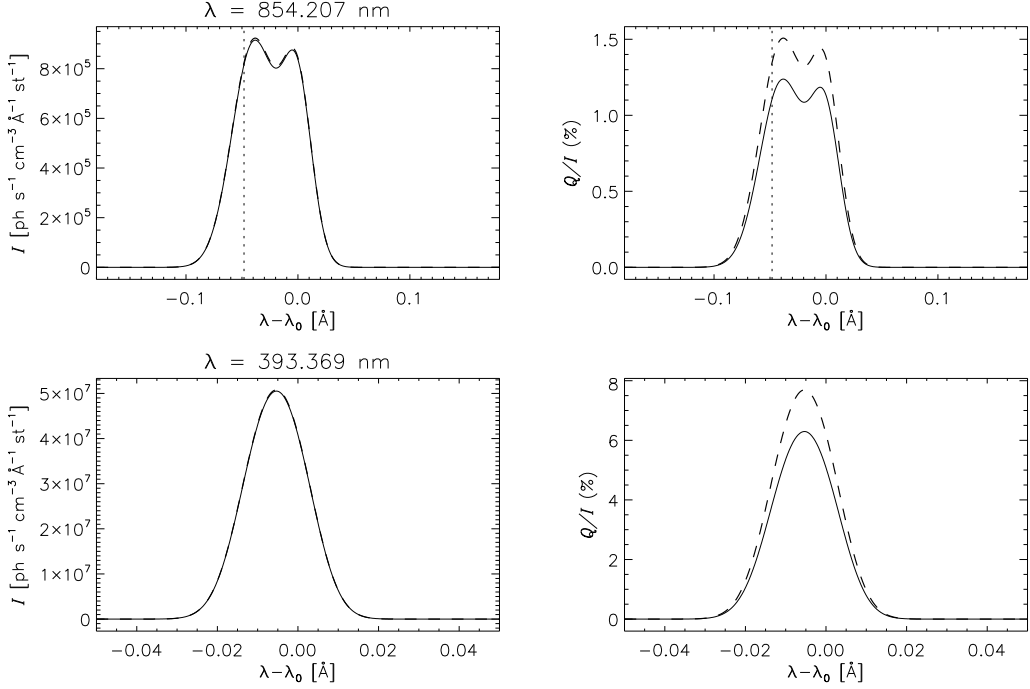}
\caption{\label{fig:plots.offres.8542}
Stokes $I$ and $Q$ emission profiles of the IR 854.2\,nm line (top panels)
and the K line at 393.4\,nm (bottom panels) of \ion{Ca}{2}, for the same
scattering model adopted for Figure~\ref{fig:plots.res.8542}. These profiles
correspond to the detuning of the incident radiation from the 
854.2\,nm wavelength resonance identified by the horizontal dashed lines 
in the contour plots of Figure~\ref{fig:plots.res.8542}. Hence, the Stokes 
$I$ profiles represent the horizontal cut of those contour plots
for the corresponding value of the detuning of the incident IR
radiation. The profiles shown with dashed curves correspond to the case 
when the atomic polarization of the metastable state of the \ion{Ca}{2}
system is neglected.}
\end{figure}

Figure~\ref{fig:plots.res.8542} shows contour plots of the scattered Stokes $I$ 
emission profiles (wavelength along the $x$ axis) of the IR line at 854.2\,nm 
(left) and the K line at 393.4\,nm (right) of \ion{Ca}{2}, as a 
function of the frequency of a monochromatic illumination (wavelength
along the $y$ axis) varying within a spectral range of 0.18\,\AA\ 
around the 854.2\,nm resonance wavelength.
Each horizontal slice of those plots thus corresponds to the scattered 
line intensity profile for the corresponding value of the detuning from 
the 854.2\,nm resonance.
%
Again, we assume no direct illumination of the
\ion{Ca}{2} K line for this modeling, and so the emitted
radiation in that line is purely produced by Raman scattering 
induced via virtual excitation of the upper state of the K line
by the IR illumination. In particular, this allows us to 
represent the scattered radiation in both lines via the contour plots 
of Figure~\ref{fig:plots.res.8542}, as a function of only one detuning
parameter.
%
The observed spectral spread of the re-emitted radiation 
along the $x$ axis is produced by Doppler redistribution,
corresponding to the plasma temperature of 1000\,K. 
%
Also for this model we assumed a zero magnetic field. In such case, the
wavelength dependence of the Stokes $Q$ polarization qualitatively resembles 
closely that of the intensity profiles (see also 
Figure~\ref{fig:plots.offres.8542}, 
and therefore we omitted showing contour plots also of Stokes $Q$. 

\begin{figure}[t!]
\centering
\includegraphics[width=\hsize]{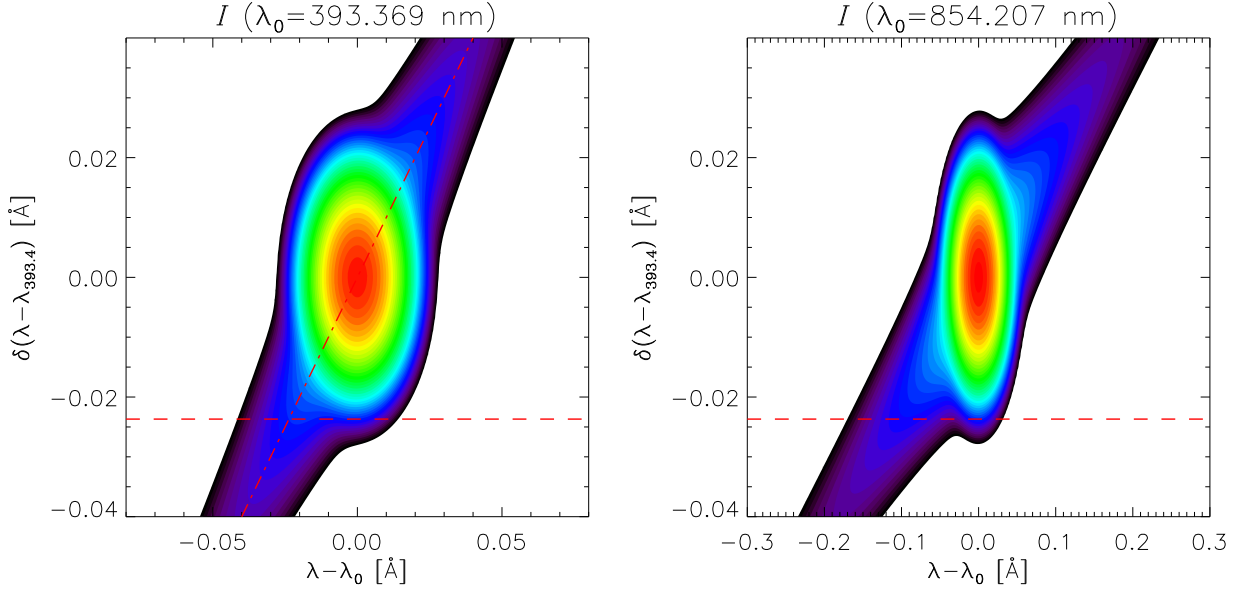}
\caption{\label{fig:plots.res.K}
Same as Figure~\ref{fig:plots.res.8542}, but for an inverted role of 
the two transitions, where the monochromatic incident radiation this
time is tuned across the spectral range of the K line centered at 
393.4\,nm, with no radiation directly exciting the IR line at 854.2\,nm.
The horizontal dashed lines locate the value of the detuning of the 
monochromatic UV incident radiation adopted for the calculation of the 
emission profiles shown in Figure~\ref{fig:plots.offres.K}.}
\end{figure}

The dash-dotted diagonal line in the left plot tracks the wavelength of 
the monochromatic IR incident radiation as it scans across the spectral 
interval around the 854.2\,nm resonance wavelength. As the detuning from 
the 854.2\,nm transition (which is read on the $y$ axis) decreases in 
absolute value, 
the signal of the scattered radiation remains initially fully coherent
with the frequency of the incident radiation, while at the same time it
increases in strength. 
In fact, for values of the detuning larger than $\sim 0.06$\,\AA\ 
in absolute value, the virtual energy levels of the upper term attained 
during the scattering process lie outside the energy band corresponding 
to the thermal line width of the upper term. Then there is no energy 
overlap leading to quantum interference between these levels and the atomic 
Hamiltonian eigen-levels of the upper term, and the scattering is purely
coherent in both the \ion{Ca}{2} K and 854.2\,nm lines.
%
This process corresponds to Rayleigh scattering of the monochromatic
incident radiation, and the frequency spread observed in the scattered 
radiation is dominated by the Doppler redistribution corresponding 
to the plasma temperature.
%

However, at about 0.06\,\AA\ from resonance, the 
signal of the non-coherent contribution at the resonance wavelength
of the 854.2\,nm line begins to appear, becoming the dominant
term of the scattered radiation in an interval of about $\pm
0.045$\,\AA\ around the resonance. The radiation emitted in the
K line, which is produced by Raman scattering towards the 
\ion{Ca}{2} ground state, is also dominated by the resonant 
contribution at 393.4\,nm approximately within the same
interval of detuning of the IR illumination, although the profiles 
of this line never become double-peaked. This can be understood if we
recall equation~(\ref{eq:important}), and it
is clearly illustrated by Figure~\ref{fig:plots.offres.8542},
which shows the Stokes $I$ and $Q$ emission profiles (solid curves) 
for a detuning of $\sim 0.05\,$\AA\ from the 854.2\,nm resonance 
wavelength (horizontal slice of the contour plots of 
Figure~\ref{fig:plots.res.8542} identified by the dashed lines). 
Overplotted on these profiles, with the dashed curves, we show the same
case where the contribution of the atomic polarization in the metastable 
state of \ion{Ca}{2} is neglected. As we see from comparing the two sets 
of profiles, the linear polarization of the scattered radiation in both 
\ion{Ca}{2} K and 854.2\,nm lines is larger in that case.

\begin{figure}[t!]
\centering
\includegraphics[width=.85\hsize]{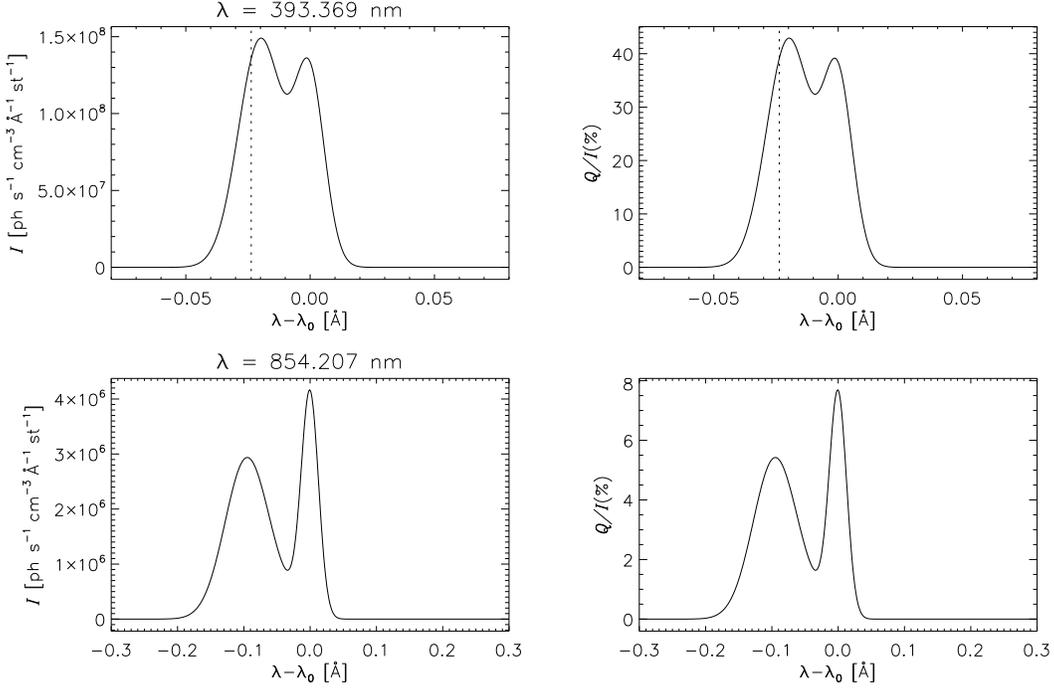}
\caption{\label{fig:plots.offres.K}
Stokes $I$ and $Q$ emission profiles of the K line at 393.4\,nm (top panels)
and the IR 854.2\,nm line (bottom panels) of \ion{Ca}{2}, for the same
scattering model adopted for Figure~\ref{fig:plots.res.8542}, but in the
case where the monochromatic incident radiation is tuned around the K line 
resonance wavelength. We note how in this case the profiles of both lines can
become visibly double-peaked, as a consequence of 
equation~(\ref{eq:important}). These profiles
correspond to the cut of the contour plots of
Figure~\ref{fig:plots.res.K} for the value of the detuning of the
monochromatic UV incident radiation identified by the horizontal dashed 
lines.}
\end{figure}

\begin{figure}[t!]
\centering
\includegraphics[width=.85\hsize]{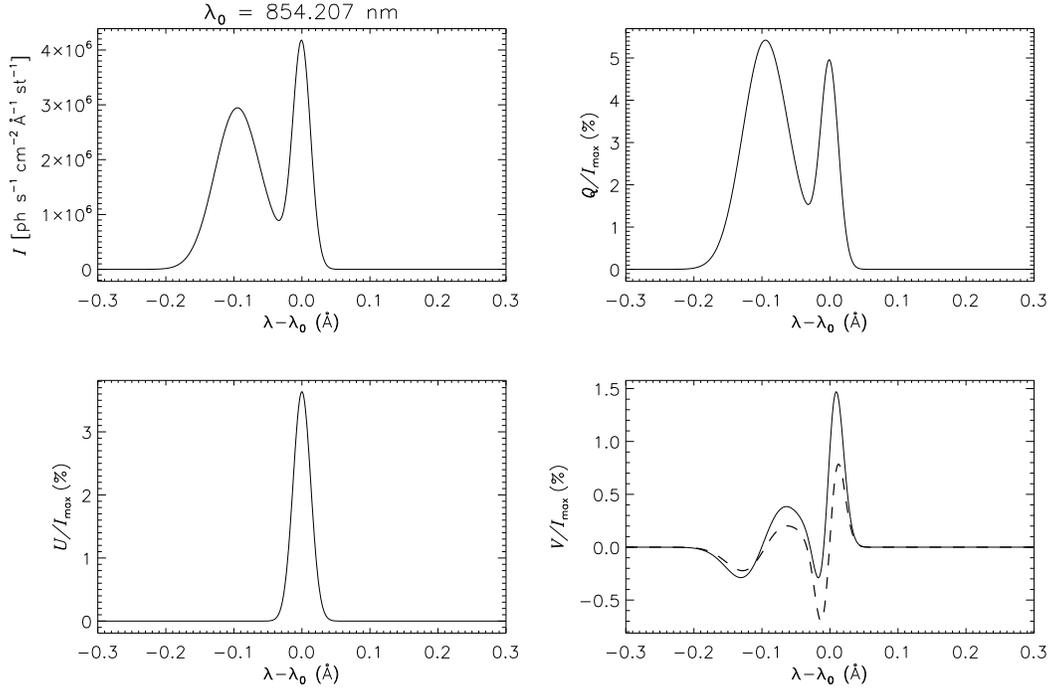}
\caption{\label{fig:plots.B}
Stokes emission profiles of the \ion{Ca}{2} K line, with the same 
detuning of the UV radiation around the K line resonance as for 
Figure~\ref{fig:plots.offres.K}, but in the case where a magnetic 
field with $B=5$\,G directed towards the observer is also present. 
We note the manifestation of the Hanle effect polarization in both 
Stokes $Q$ and $U$, in the resonant core of the line.
The dashed curve in the panel for Stokes $V$ represents the 
weak-field approximation to the circular polarization signal, 
computed through the first derivative of Stokes $I$, assuming the
$LS$-coupling value for the effective Land\'e factor of the line, 
$g_{\rm eff}\approx 1.1$.}
\end{figure}

Figures~\ref{fig:plots.res.K} and \ref{fig:plots.offres.K} are analogous
to Figures~\ref{fig:plots.res.8542} and \ref{fig:plots.offres.8542},
for the case where the monochromatic incident radiation is tuned 
across the resonance wavelength of the K line at 393.4\,nm instead, 
with no radiation directly exciting the IR transition at 854.2\,nm. 
As evidenced by Figure~\ref{fig:plots.offres.K}, in this case the 
profiles of both lines can become double-peaked (for values of the 
detuning from resonance larger than $\sim 0.02\,$\AA\ in absolute value), 
in virtue of the separation between the coherent and resonant 
contributions satisfying equation~(\ref{eq:important}).

When a magnetic field is present, the profiles evidently become more 
complicated, although some general conclusions can be drawn from the example 
presented in Figure~\ref{fig:plots.B}. This shows the full Stokes 
profiles of the \ion{Ca}{2} IR line at 854.2\,nm under the same 
illumination conditions of Figure~\ref{fig:plots.offres.K}, but with 
the addition of a magnetic field of 5\,G directed towards the observer 
(i.e., normal to the direction of the incident radiation).
Because the magnetic field strength is comparable to
the Hanle critical field for the upper level ${}^2P_{3/2}$ 
of the \ion{Ca}{2} 854.2\,nm and K lines
($B_{\rm Hanle}\approx \epsilon_u/(0.8794\times10^7\,g_u) 
\approx 6.7$\,G, where $g_u\approx 1.333$ is the Land\'e factor of 
the ${}^2P_{3/2}$ level, and $\epsilon_u$ is calculated via
equation~(\ref{eq:widths})), one expects that the Stokes profiles will show 
evidence of Hanle-effect depolarization. 
Indeed, comparing the bottom panels of Figure~\ref{fig:plots.offres.K} 
with the top panels of Figure~\ref{fig:plots.B}, 
we see that the intensity profile is practically unaffected by the 
presence of a weak magnetic field, whereas the resonant component 
of Stokes $Q$ shows a depolarization with respect to the zero-field 
case, and accordingly a signal in Stokes $U$ appears.
It is important to observe that the Stokes $U$ signal manifests
itself strictly in the resonant core of the line, which is also where the 
depolarization of Stokes $Q$ occurs.
This is to be expected, since the Hanle effect is a manifestation of the
relaxation of atomic coherence as the energy degeneracy of the 
atomic Hamiltonian eigen-levels is lifted by the applied magnetic field.


\begin{figure}[t!]
\centering
\includegraphics[width=.85\hsize]{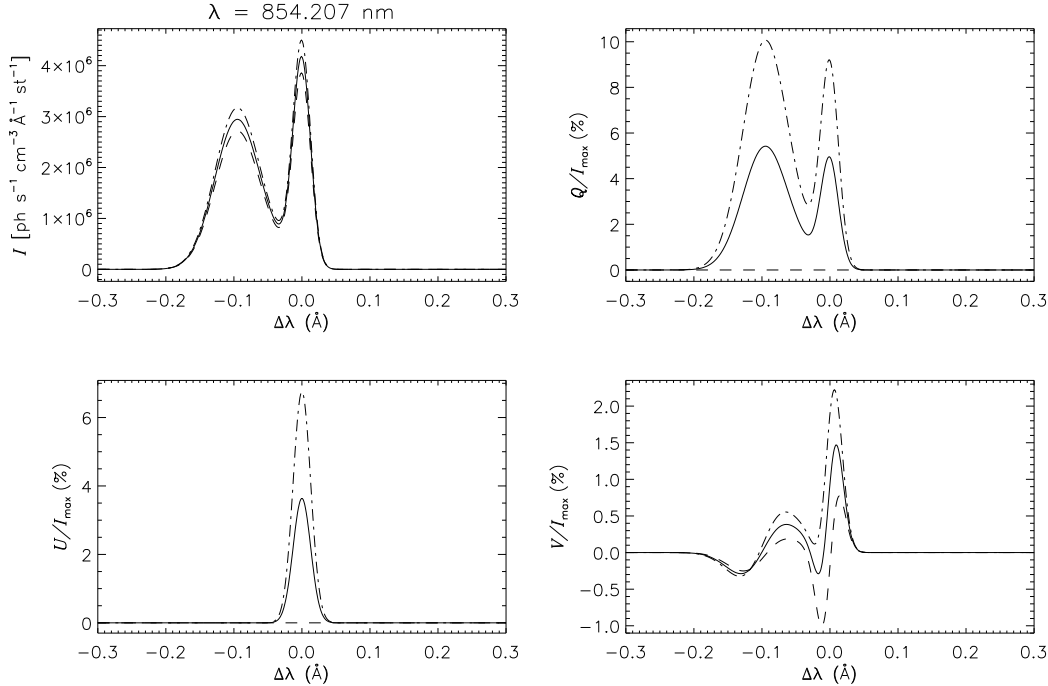}
\caption{\label{fig:Ca_8542.B+Q}
Stokes profiles of the \ion{Ca}{2} line at 854.2\,nm, for the same 
scattering configuration and magnetic model of 
Figure~\ref{fig:plots.B}, and different conditions of polarization 
of the monochromatic UV incident radiation in the proximity of the 
\ion{Ca}{2} K line: 
unpolarized (solid curves; identical to the case of 
Figure~\ref{fig:plots.B}); fully linearly polarized along the LOS (dashed 
curves); fully linearly polarized perpendicularly to the LOS 
(dash-dotted curves).}
\end{figure}

We also note that the shape of Stokes $V$ is not quite reproduced 
by the weak-field approximation of the circular polarization 
signal (dashed curve in the bottom-right panel of 
Figure~\ref{fig:plots.B}), despite the very small strength of the
applied field, and the fact that all polarization effects due to 
level interference induced by the magnetic field, 
such as Stokes-$V$ asymmetries associated with atomic orientation,
are completely negligible in this case, because of the relatively 
large fine structure separation between the $J=1/2,3/2$ levels of 
the upper term.
Therefore, the net circular polarization observed
in Figure~\ref{fig:plots.B} is a manifestation of the effects of 
partial redistribution on the circular polarization of the 
scattered light.

All previous examples were calculated assuming an unpolarized 
beam of incident radiation.
Figure~\ref{fig:Ca_8542.B+Q} shows instead the Stokes profiles of the 
\ion{Ca}{2} 854.2\,nm line for the same scattering configuration and magnetic 
model as in the example of Figure~\ref{fig:plots.B},
but under different conditions of polarization of the incident UV
radiation.
More precisely, the atomic system is still prepared assuming the same 
flat and unpolarized radiation field as in all previous examples, and only 
the incident beam of radiation used in equation~(\ref{eq:RT.J}) is now
assumed to be linearly polarized.
The solid curves represent the case of unpolarized incident 
radiation as a reference, which are identical to those shown in 
Figure~\ref{fig:plots.B}. The other curves represent instead the cases 
of fully linearly polarized radiation with $Q/I=+1$ (polarization along 
the LOS; dashed curves) and $Q/I=-1$ (polarization perpendicular to 
the LOS; dash-dotted curves).
When the incident beam of radiation is linearly polarized along the LOS, 
the linear polarization of the scattered radiation is
practically completely suppressed (in fact, by approximately six 
orders of magnitude compared to the unpolarized case),
as it can be expected also on the basis of 
simple classical arguments \citep{MZ34}.

As a concluding remark, we want to point out that, while the atomic
polarization produced by the anisotropy of the incident radiation is fully 
accounted for in the modeling examples presented in this work, all 
those examples are realized under physical conditions where the atomic
coherence in the lower terms is practically negligible. This is
either because the magnetic field is absent (Figures~\ref{fig:Lyb-Ha} to
\ref{fig:plots.offres.K}), or because the magnetic strength is above
the critical value for the Hanle effect of the lower term 
(Figures~\ref{fig:plots.B} and \ref{fig:Ca_8542.B+Q}), yet small enough 
not to induce any $J$-$J'$ 
level interference via the Paschen-Back effect. 
The realization of this range of physical conditions allows us to adopt 
a reduced form of the redistribution function for the polarized atom
introduced in \citeauthor{Ca14} (\citeyear{Ca14}), 
which corresponds to a direct generalization to the three-term atom 
of the $R_{\rm II}$ function adopted elsewhere in the PRD literature 
(e.g., \citealt{BT14}).

\begin{acknowledgments}
We thank Tanaus\'u del Pino Alem\'an (Instituto de Astrof\'{\i}sica 
de Canarias; IAC) for helpful discussions during the initial testing of 
the code with which the results of this paper were produced. We also
thank Rebecca Centeno Elliot (HAO) and Javier Trujillo Bueno (IAC)
for reading the manuscript and for providing helpful comments and 
suggestions. RMS acknowledges partial support by the Spanish Ministry of 
Economy and Competitiveness (project AYA2010-18029), and the support of 
the HAO Visitor Program, which made this work possible.

We dedicate this work to the memory of David G.\ Hummer (1934-2015), 
who greatly contributed to the field of frequency redistribution.
David, who for many years of his active career was a vital collaborator 
of HAO and NCAR, as witnessed by his many works published together 
with HAO scientists, passed away on December 17, while this paper was 
receiving the final strokes.  
\end{acknowledgments}

\begin{appendix}

\section{Redistribution function of the $\Lambda$-type three-term 
atom in the laboratory frame}

In the case of infinitely sharp lower levels (s.l.l.), and assuming that the
initial term of the $\Lambda$-type transition is \emph{non-coherent} (i.e.,
$\rho_{ll'}=\delta_{ll'}\rho_{ll}$), the redistribution function for the
$\Lambda$-type three-term polarized atom, expressed in the laboratory
reference frame, is given by
\begin{eqnarray} \label{eq:R.last}
R(\Omega_u,\Omega_{u'};\Omega_l,\Omega_{f};
	\hat{\omega}_{k},\hat{\omega}_{k'};\Theta)_{\rm s.l.l.} 
&=& \frac{2\pi}{\Delta^2\,S\xi_l\xi_f}
\exp\biggl[-\frac{(\hat\omega_k-\hat\omega_{k'}+\omega_{lf})^2}
	{\Delta^2}\biggr] \\
&&\kern -3cm {}\times \biggl[
W\biggl(\frac{\kappa^+ v_{ul}+\kappa^- w_{uf}}{S\xi_l\xi_f},
	\frac{a_u}{S\xi_l\xi_f} \biggr) +
\overline{W}\biggl(\frac{\kappa^+ v_{u'l}+\kappa^- w_{u'f}}{S\xi_l\xi_f},
	\frac{a_{u'}}{S\xi_l\xi_f} \biggr) \biggr]\;, \nonumber
\end{eqnarray}
where
\begin{equation} \label{eq:W}
W(v,a)=\frac{1}{\pi}\int_{-\infty}^{+\infty} dp\;
\frac{{\rm e}^{-p^2}}{a+{\rm i}(p-v)}=H(v, a)+{\rm i}\,L(v, a)\;,
\end{equation}
with $H(v,a)$ and $L(v,a)$ being respectively the Voigt and Faraday-Voigt 
functions. The full derivation of the above result is contained
in an upcoming paper (R.\ Casini \& R.\ Manso Sainz 2016; in preparation).

For each scattering event, $\Theta$ is the angle between the propagation 
directions of the incoming and outgoing photons.
We introduced accordingly the associated quantities
\begin{equation}
C=\cos\Theta\;,\qquad S=\sin\Theta\;.
\end{equation}
Next we defined
\begin{equation} \label{eq:avg_Doppler}
\Delta=(\Delta_{ul}^2+\Delta_{uf}^2 - 2C\Delta_{ul}\Delta_{uf})^{1/2}\;,
\end{equation}
\begin{equation} \label{eq:xi}
\xi_l=\Delta_{ul}/\Delta\;,\qquad
\xi_f=\Delta_{uf}/\Delta\;,
\end{equation}
where $\Delta_{mn}$ is the Doppler width of the transition between the 
atomic states $m$ and $n$, with Bohr frequency $\omega_{mn}$. 
We introduced the normalized frequency variables
\begin{equation} \label{eq:variables.1}
v_{mn}=(\hat\omega_k-\omega_{mn})/\Delta\;,\qquad
w_{mn}=(\hat\omega_{k'}-\omega_{mn})/\Delta\;,
\end{equation}
where the incoming and outgoing radiation frequencies, $\hat\omega_k$ and 
$\hat\omega_{k'}$, are expressed in the laboratory frame of reference. 
We also introduced normalized damping parameters associated with the inverse
lifetimes of the transition levels, using the same ``reduced'' Doppler width
of equation~(\ref{eq:avg_Doppler}),
\begin{equation}
\label{eq:variables.2}
a_m=\epsilon_m/\Delta\;.
\end{equation}
Finally, we introduced the transitions' ``weights''
\begin{equation} \label{eq:chi}
\kappa^\pm={\textstyle\frac{1}{2}}\bigl[1\pm(\xi_f^2-\xi_l^2)\bigr]\;.
\end{equation}
\end{appendix}

\end{document}